\documentstyle[11pt,epsf,Axodraw]{article}
\input epsf.tex
\newcommand{\be}{\begin{equation}}
\newcommand{\ee}{\end{equation}}
\newcommand{\bea}{\begin{eqnarray}}
\newcommand{\eea}{\end{eqnarray}}

\begin{document}
\title{Photon Neutrino Scattering in Non-Commutative Space}
\author{M. Haghighat\thanks{email: mansour@cc.iut.ac.ir},
 \ \ M. M. Ettefaghi,\ \ \ M. Zeinali\\ \\
{\it Department of  Physics, Isfahan University of Technology (IUT)}\\
{\it Isfahan 84156,  Iran,} \\}

\date{}
\maketitle

\begin{abstract}
We extend the non-commutative standard model based on the minimal
$SU(3)\times SU(2)\times U(1)$ gauge group to include the
interaction of photon with neutrino.  We show that, in the gauge
invariant manner, only the right handed neutrino can directly
couple to the photon. Consequently, we obtain the Feynman rule
for the $\gamma\nu\bar\nu$-vertex which does not exist in the
minimal extension of non-commutative standard model (mNCSM).  We
calculate the amplitude for $\gamma\nu\rightarrow\gamma\nu$ in
both the nonminimal non-commutative standard model (nmNCSM) and
the extended version of mNCSM.  The obtained cross section grows
in the center of mass frame, respectively, as
$(\theta_{NC})^2{M}_Z^{-4}E^6$ and $(\theta_{NC})^4E^6$ which can
exceed the cross section for
$\gamma\nu\rightarrow\gamma\gamma\nu$ and
$\gamma\nu\rightarrow\gamma\nu$ in the high energy limit in the
commutative space.
\end{abstract}
\section{Introduction}
High energy photon and neutrino and their scattering channels
based on the standard model are currently of interest to many
authors in astrophysics and cosmology
\cite{ast-pn}-\cite{dicus1}. In the low energy limit the elastic
photon-neutrino scattering is strongly suppressed by Yang's
theorem in the lowest order \cite{yang}. Meanwhile, the inelastic
scattering of photon-neutrino such as
$\gamma\nu\rightarrow\gamma\gamma\nu$ and its crossed processes
are not subject to this suppression i.e.
$\sigma_{\gamma\nu\rightarrow\gamma\gamma\nu}(1MeV)\sim 10^{-52}
cm^2 $ in comparison with
$\sigma_{\gamma\nu\rightarrow\gamma\nu}(1MeV)\sim 10^{-65} cm^2$
\cite{dicus}.  Nevertheless, in the high energy limit it is shown
that \cite{abbas}
\begin{equation}
 \sigma_{\gamma\nu\rightarrow\gamma\nu}=6.7\times 10^{-33}(\frac{E}{m_e})^6
 \,\,\,pb,\label{elastic pn}
 \end{equation}
 while \cite{dicus1}
\begin{equation}
 \sigma_{\gamma\nu\rightarrow\gamma\gamma\nu}=1.74 \times 10^{-16}(\frac{E}{m_e})^2
  \,\,\,pb,\label{inelastic pn}
 \end{equation}
in which the photon energy, $E$, in the center of mass frame
satisfies $m_e\ll E\ll M_W$.  Obviously, with increasing $E$ the
elastic cross section exceeds the inelastic one and it can be
easily seen that the crossover occurs at $E\sim 7GeV$.  In the
high energy limit the non-commutativity effects seem to be
significant and therefore the new interactions of photon and
neutrino in the non-commutative space and time can be potentially
important to astrophysics. However, non-commutative field theory
and its phenomenological aspects have been recently considered by
many authors \cite{SW}-\cite{sun}. Such theories are mostly
characterized on a non-commutative space-time with  the
non-commutativity parameter $\theta_{\mu\nu}$. In the canonical
version of the non-commutative space-time one has
\begin{equation}
\theta^{\mu\nu}=-i\left[\hat{x}^\mu,\hat{x}^\nu\right],
\end{equation}
where a hat indicates a non-commutative coordinate and
$\theta_{\mu\nu}$ is a real, constant and anti-symmetric matrix.
The action for field theories on non-commutative spaces is then
obtained by using the Weyl-Moyal correspondence;  Accordingly, the
usual product of fields should be replaced by the star product:
\begin{equation}
 f\star
g(x,\theta)=f(x,\theta)\exp(\frac{i}{2}\overleftarrow{\partial}_\mu
\theta^{\mu\nu}\overrightarrow {\partial}_\nu)g(x,\theta).
\label{a3}
\end{equation}
In replacing the ordinary product with the star product there is
an ambiguity in transcribing $gA\psi$, where $g$, $A$ and $\psi$
are coupling constant,  gauge and  particle fields, respectively,
into non-commutative form that is: $gA\star\psi$, $g\psi\star A$
or $g_1A\star\psi + g_2\psi\star A$. In the commutative limit all
the terms can be reduced to the same term while for the neutral
particles, for example in QED, the third term in the
non-commutative limit is essentially different from the other
two.  In fact this can bring about direct interaction of photon
and neutral particles.

In section 2 we give a brief review on the direct interaction of
neutral particles with photon in the non-commutative QED and
subsequently extend the non-commutative standard model (NCSM)
based on $SU(3)_c\times SU(2)_L\times U(1)_Y$ gauge group to
incorporate the direct interaction of photon with neutrino. In
section 3 we explore the photon-neutrino elastic scattering in the
extended minimal NCSM as well as  the non-minimal NCSM at the
lowest order. Finally, we compare our results with the results on
the photon-neutrino scattering given in the literature.
 \section{Non-commutative standard model}
In the frame work of NCQED it is shown that the neutral particles
interact with photons if they transform under U(1) in a similar
way as in the adjoint representation of a non-Abelian gauge
theory.  In fact, for this purpose $eA\star\psi -e\psi\star A$
should be added to ordinary derivative to construct the covariant
derivative \cite{pn}-\cite{pnqed}. In the limit of
$\theta\rightarrow 0$, we have
\begin{equation}
eA\star\psi -e\psi\star A=0+\cal{O}(\theta),
\end{equation}
therefore the covariant derivative to the lowest order can be
obtained as follows
\begin{equation}
\hat{D_{\mu}}\hat{\psi}={\partial}_\mu\hat{\psi}+e{\theta}^{\nu\rho}{\partial}_\nu\hat{A}_\mu
{\partial}_\rho\hat{\psi}.
\end{equation}
The fields themselves in the non-commutative space can be
expanded by the Seiberg-Witten (SW)  map \cite{SW} up to the
lowest order as
\begin{equation}
\hat{\psi}=\psi +
e{\theta}^{\nu\rho}A_\rho{\partial}_\nu\psi,\label{psi}
\end{equation}
\begin{equation}
\hat{A_\mu}=A_\mu + e{\theta}^{\nu\rho}A_\rho[{\partial}_\nu
A_\mu-\frac{1}{2}{\partial}_\mu A_\nu].\label{A}
\end{equation}
Therefore the interaction term in terms of commutative fields is
\begin{equation}
-\frac{e}{2}F_{\mu\nu}(i{\theta}^{\mu\nu\rho}{\partial}_\rho)\psi,
\end{equation}
where $F_{\mu\nu}={\partial}_\mu A_\nu-{\partial}_\nu A_\mu$ and
\begin{equation}
{\theta}^{\mu\nu\rho}={\theta}^{\mu\nu}\gamma^\rho+{\theta}^{\nu\rho}\gamma^\mu
+{\theta}^{\rho\mu}\gamma^\nu.
\end{equation}
It should be noted that for neutrino as a neutral particle in the
NCQED, as well as QED, in contrast with the standard model, there
is not any constraint on the mass or even the chirality of the
neutrino. In the standard model, neutrino is massless and only the
left handed one has weak interaction while the right handed
neutrino, if existing, has an expectator role in all reactions.
However, there are two approaches to construct the standard model
in the non commutative space.  In the minimal extension the gauge
group is $SU(3)_c\times SU(2)_L\times U(1)_Y$ in which the number
of gauge fields, couplings and particles are the same as the
ordinary one \cite{NCSM}.  Although in this extension new
interactions will appear due to the star product and the SW map,
the photon-neutrino vertex is absent.  In the second approach the
gauge group is $U(3)\times U(2)\times U(1)$ which is reduced to
$SU(3)_c\times SU(2)_L\times U(1)_Y$ by an appropriate symmetry
breaking \cite{nNCSM}. However, in the latter approach, besides
many new interaction like the former one, photon can interact
with the left handed neutrino.

To introduce the neutrino-photon interaction in the minimal NCSM,
one can define the adjoint representation in the covariant
derivative for the neutral particle as is done in the NCQED.  The
main difference in the SM is $U(1)_Y$ instead of  $U(1)_{EM}$.
Therefore neutral hyper charge particle can only couple to the
hyper gauge field in a gauge invariant manner.  The only particle
with zero hyper charge in the SM is the right handed neutrino
therefore the covariant derivative for this particle can be
written as follows
\begin{equation}
\hat{D_{\mu}}\hat{\psi}_{\nu_R}={\partial}_\mu\hat{\psi}_{\nu_R}+
e{\theta}^{\nu\rho}{\partial}_\nu\hat{B}_\mu
{\partial}_\rho\hat{\psi}_{\nu_R},
\end{equation}
in which $\hat{\psi}_{\nu_R}$ and $\hat{B}$, respectively, denote
the NC-fields of the right handed neutrino and the hyper charge
with their own expansion in the NC-space as are given in
Eqs.(\ref{psi}-\ref{A}). Consequently, Lagrangian density for the
right handed neutrino part of NCSM can be written as follows
\begin{eqnarray}
&{\cal{L}}_{\nu_R}&=i{\bar{\psi}}\partial\!\!\!\!/\psi +
ie{\theta}^{\mu\nu}[\partial_\mu{\bar{\psi}}B_\nu\gamma^\rho({\partial}_\rho\psi)
\nonumber\\& &
-\partial_\rho{\bar{\psi}}B_\nu\gamma^\rho({\partial}_\mu\psi) +
{\bar{\psi}}( \partial_\mu B_\rho
)\gamma^\rho({\partial}_\nu\psi)]\label{Lagden},
\end{eqnarray}
where $B$ in terms of the photon and the $Z$-gauge boson fields is
\begin{equation}
B= \cos\theta_W A-\sin\theta_W Z.
\end{equation}
Therefore the Feynman rules for $\gamma\nu\bar\nu$ and
$Z\nu\bar\nu$  vertices can be obtained from the Lagrangian
(\ref{Lagden}) as:
\begin{equation}
\Gamma^\mu_{\gamma\nu\bar\nu}=i{\frac
{e}{2}}cos\theta_W(1+\gamma_5)(\theta^{\mu\nu}k_\nu q \!\!\!/ +
\theta^{\rho\mu}q_\rho k \!\!\!/ + \theta^{\nu\rho}k_\nu
q_\rho\gamma^\mu),\label{gnn}
\end{equation}
and
\begin{equation}
\Gamma^\mu_{Z\nu\bar\nu}=-i{\frac
{e}{2}}sin\theta_W(1+\gamma_5)(\theta^{\mu\nu}k_\nu q\!\!\!/ +
\theta^{\rho\mu}q_\rho k\!\!\!/ + \theta^{\nu\rho}k_\nu
q_\rho\gamma^\mu).\label{znn}
\end{equation}
It should be noted here that in the minimal extension of the
standard model to the non-commutative space-time (mNCSM) there is
not any $\gamma\nu\bar\nu$-vertex while the $Z\nu\bar\nu$-vertex
has already existed for the left handed neutrino therefore
$\Gamma^\mu_{Z\nu\bar\nu}$ for the right handed neutrino can be
considered as a correction to the same vertex in the mNCSM.  Since
the other particles in the SM, even the left handed neutrino, all
have nonzero hyper charge, the remaining parts of the SM in the
noncommutative space do not change.
 \section{Photon-neutrino interaction in NCSM}
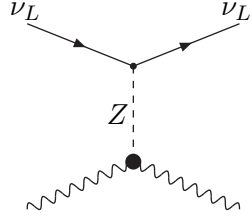
\begin{figure}[tp]
\begin{picture}(300,50)(0,0)
\ArrowLine(200,35)(240,19)\ArrowLine(240,19)(280,35)
\DashLine(240,-17)(240,19){3}
\Photon(200,-35)(240,-17){2}{7.5}\Photon(240,-17)(280,-35){2}{7.5}
\Vertex(240,19){1.3}\Vertex(240,-17){3} \Text(200,40)[c]{$\nu_L$}
\Text(280,40)[c]{$\nu_L$}\Text(235,1)[c]{$Z$}
\end{picture}
\\\\\\
\caption{Feynman diagram for the process
$\gamma\nu\rightarrow\gamma\nu$ in nmNCSM at the order $\theta$.
The bold dot represents  the non-commutative vertex
${\Gamma}^{\mu\nu\rho}$.}
\end{figure}
In the minimal extension of the standard model to the
non-commutative space-time due to the different choices for
representations of the gauge group the trace in the kinetic terms
for gauge bosons is not unique.  In fact the freedom in the
choice of the traces can be used to construct a new version of
the NCSM which is called nmNCSM.  Neutral triple-gauge boson
vertices such as $\gamma \gamma\gamma$ and $Z\gamma\gamma$ in
contrast to the mNCSM as well as SM can arise within the framework
of the nmNCSM.  These vertices can be extracted from the
Lagrangian of nmNCSM which are given in \cite{NCSM} as follows
\begin{equation}
{\cal L}_{\gamma \gamma\gamma}={\frac {e}{4}}sin2\theta_WK_{\gamma
\gamma\gamma}\theta^{\rho\sigma}A^{\mu\nu}(
A_{\mu\nu}A_{\rho\sigma}-4A_{\mu\rho}A_{\nu\sigma}),
\end{equation}
\begin{eqnarray}
&{\cal L}_{Z \gamma\gamma}&={\frac {e}{4}}sin2\theta_WK_{Z
\gamma\gamma}\theta^{\rho\sigma}[2Z^{\mu\nu}(
2A_{\mu\rho}A_{\nu\sigma}-A_{\mu\nu}A_{\rho\sigma})\nonumber\\& &
+8Z_{\mu\rho}A^{\mu\nu}A_{\nu\sigma}
-Z_{\rho\sigma}A_{\mu\nu}A^{\mu\nu}],\label{zgg}
\end{eqnarray}
and
\begin{equation}
{\cal L}_{ZZ\gamma}={\cal L}_{Z \gamma\gamma}(A_\mu
\leftrightarrow Z_\mu),
\end{equation}
\begin{equation}
{\cal L}_{ZZZ}={\cal L}_{\gamma\gamma\gamma}(A_\mu \rightarrow
Z_\mu),
\end{equation}
where
\begin{equation}
A_{\mu\nu}={\partial}_\mu A_\nu-{\partial}_\nu A_\mu,
\end{equation}
\begin{equation}
Z_{\mu\nu}={\partial}_\mu Z_\nu-{\partial}_\nu Z_\mu.
\end{equation}
The constants $K_{\gamma \gamma\gamma}$, $K_{Z \gamma\gamma}$ and
so on are functions of the coupling constants of the
non-commutative electroweak sector up to the first order of
$\theta$.  They can be obtained by matching the NCSM action with
the SM action and their allowed range of values is given in
\cite{kzNCSM}. However, up to the first order of $\theta$ in the
nmNCSM, there is a Feynman diagram which is shown in Fig.(1). The
Feynman rule for the $Z \gamma\gamma $ vertex in the nmNCSM can
be easily derived from the Lagrangian of Eq.(\ref{zgg}) as follows\\
\begin{picture}(300,25)(0,0)
\unitlength=1mm \DashLine(340,-17)(340,19){3}
\Photon(300,-35)(340,-17){2}{7.5}\Photon(340,-17)(380,-35){2}{7.5}
\Vertex(340,-17){3}\Text(57,-7)[c]{${\Gamma}^{\mu\nu\rho}=-2esin\theta_WK_{Z
\gamma\gamma}\Theta((\mu,k_1),(\nu, k_2), (\rho,k_3)),$} \nonumber
\end{picture}
\vspace{1.5cm}\\
in which $K_{Z \gamma\gamma}$ is the strength of the $Z
\gamma\gamma$ triple-gauge bosons and
\begin{eqnarray}
&\Theta((\mu,k_1),(\nu, k_2),
(\rho,k_3))=-\theta^{\mu\nu}(k^\rho_1(k_2.k_3)-k^\rho_2(k_1.k_3))&
 \nonumber\\& + \theta^{\mu\alpha}k_{1\alpha}(g^{\nu\rho}(k_2.k_3)-k^{\rho}_2k^{\nu}_3)
  -\theta^{\nu\alpha}k_{1\alpha}(g^{\rho\mu}(k_2.k_3)-k^{\rho}_2k^{\mu}_3)&
 \nonumber\\&  -\theta^{\rho\alpha}k_{1\alpha}(g^{\mu\nu}(k_2.k_3)-k^{\mu}_2k^{\nu}_3)
  +k_1.\theta
  .k_2(k^\mu_3g^{\nu\rho}-k^\nu_3g^{\rho\mu})&\nonumber\\&
 + cycl.\,\, permut.\,\, of(\mu_i,k_i),&
\label{m1}
\end{eqnarray}
where $\mu_1=\mu$, $\mu_2=\nu$ and $\mu_3=\rho$.
 Therefore, the invariant amplitude for the reaction
\begin{equation}
\gamma(k_1,\varepsilon_\mu)+\nu(k_3)\rightarrow
\gamma(p_2,\varepsilon_\rho)+\nu(p_1)
\end{equation}
can be easily written as
\begin{eqnarray}
&-i{\cal M}&=\varepsilon_\mu(k_1)\varepsilon_\rho(p_2)T^{\mu\rho}\nonumber\\
& & =\varepsilon_\mu(k_1)\varepsilon_\rho(p_2)\overline{u}(p_1)
\frac{-ig}{2\cos\theta_W}\gamma_\nu\frac{1}{2}(1-\gamma^5)\nonumber\\
& &u(k_3)
\frac{i(-2e\sin2\theta_Wk_{Z\gamma\gamma})}{M_Z^2-k^2_2}\Theta^{\mu\nu\rho},\label{m1}
\end{eqnarray}
where, after some algebra $\Theta^{\mu\nu\rho}$ in the center of
mass frame, can be obtained as:
\begin{eqnarray}
&\Theta^{\mu\nu\rho}\
=&\Bigg\{2(k_1.p_2)\theta^{\mu\alpha}p_{2_\alpha}g^{\nu\rho}-p_2^\rho
p_2^\nu\theta^{\mu\alpha}k_{2_\alpha}+k_1^\rho
k_1^\mu\theta^{\nu\alpha}p_{2_\alpha}-p_2^\mu p_2^\rho
\theta^{\nu\alpha}k_{1_\alpha}-k_1^\nu
k_1^\mu\theta^{\rho\alpha}k_{2_\alpha} \nonumber\\& &
-2(k_1.p_2)\theta^{\rho\alpha}k_{1_\alpha}g^{\mu\nu}-(k_1.p_2)\theta^{\mu\nu}p_2^\rho
-(k_1.p_2)\theta^{\nu\rho}k_1^\mu+2(k_1.p_2)\theta^{\rho\mu}k_1^\nu\nonumber\\
& &+(k_1.\theta.p_2)(k_1^\mu-2p_2^\mu)g^{\nu\rho}+2
(k_1.\theta.p_2)k_1^\nu
g^{\rho\mu}+(k_1.\theta.p_2)(p_2^\rho-2k_1^\rho)g^{\mu\nu}\Bigg\},
\end{eqnarray}
and as a natural consequence of gauge symmetry one can easily
show that $T^{\mu\rho}$ satisfies the Ward identity as
\begin{equation}
k_{1\mu}T^{\mu\rho}=p_{2\rho}T^{\mu\rho}=0.
\end{equation}
It therefore follows that if $E\ll M_Z$ then, after a little
algebra, the spin-averaged amplitude is
\begin{eqnarray}
&\overline{\mid {\cal M}\mid}^2\
=&\Bigg(\frac{4\pi\alpha}{M_Z^2}\Bigg)^2\mid
k_{Z\gamma\gamma}\mid^2\times2^5\Bigg\{(k_1.p_2)^3\Big(p_2.\theta.\theta.p_2+k_1.\theta.\theta.k_1\Big)\nonumber\\&&
+(k_1.p_2)^2\Bigg(p_1.\theta.\theta.p_2(k_1.k_3)+k_3.\theta.\theta.p_2(p_1.k_1)-k_1.\theta.\theta.p_2(k_1.p_2)\nonumber\\&&+
p_1.\theta.\theta.k_1(k_1.k_3)+k_3.\theta.\theta.k_1(p_1.k_1)-k_1.\theta.\theta.k_1(k_1.p_2)\Bigg)\nonumber\\&&
-(k_1.p_2)^2\Bigg((p_1.\theta.k_1)(k_3.\theta.p_2)+k_1.\theta.\theta.p_2(k_1.p_2)\Bigg)\nonumber\\&&
+(k_1.p_1)(k_1.k_3)\Bigg((k_1.p_2)^2\mid\!\!\overrightarrow{\theta}\!\!\mid^2+2(k_1.\theta.p_2)^2\Bigg)\Bigg\},
\end{eqnarray}
thus by doing some manipulation the total cross section for
$\gamma\nu\rightarrow\gamma\nu$ in nmNCSM results in
\begin{eqnarray}
\sigma\cong11.5\mid
k_{Z\gamma\gamma}\mid^2\frac{\alpha^2E^6}{\Lambda^4M_Z^4}\label{cross-nm},
\end{eqnarray}
where the scale of non-commutativity $\Lambda$ is defined as
\begin{eqnarray}
\Lambda=\frac{1}{\sqrt{\mid\!\!\overrightarrow{\theta}\!\!\mid^2}}\label{lambda}.
\end{eqnarray}
\begin{figure}[tp]
\begin{picture}(300,50)(0,0)
\unitlength=1mm
\Photon(60,19)(100,1){2}{7.5}\ArrowLine(100,1)(140,1)\ArrowLine(60,-17)(100,1)
\Photon(140,1)(180,19){2}{7.5}\ArrowLine(140,1)(180,-17)
\Vertex(100,1){3}\Vertex(140,1){3} \Text(19,-7)[c]{$\nu_R$}
\Text(67,-7)[c]{$\nu_R$}\Text(42,-15)[c]{(1)}
\Photon(230,35)(270,19){2}{7.5}\ArrowLine(270,19)(310,35)
\ArrowLine(270,-17)(270,19)
\ArrowLine(230,-35)(270,-17)\Photon(270,-17)(310,-35){2}{7.5}
\Vertex(270,19){3}\Vertex(270,-17){3} \Text(80,-15)[c]{$\nu_R$}
\Text(110,14)[c]{$\nu_R$}\Text(95,-15)[c]{(2)}
\nonumber
\end{picture}
\\\\\\
\caption{Feynman diagrams for the process
$\gamma\nu\rightarrow\gamma\nu$ in mNCSM.  The bold dot represents
the non-commutative vertex $\Gamma^\mu_{\gamma\nu\bar\nu}$.}
\end{figure}
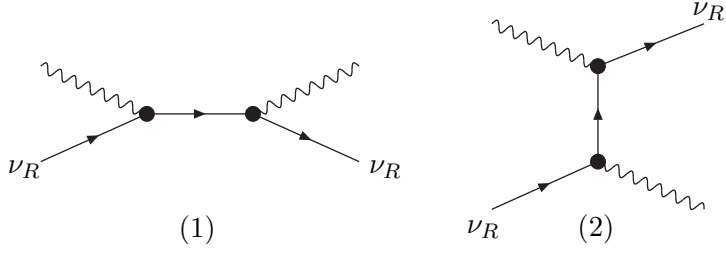
The constant $K_{Z \gamma\gamma}$ varies in the range $-0.3\leq
K_{Z \gamma\gamma}\leq 0.1$ and for $K_{Z \gamma\gamma}\sim 0.1$
the cross section varies in the range $10^{-42}$-$10^{-46}$ $cm^2$
for $\Lambda\sim 100-1000 GeV$ and $E\sim 0.1M_Z$ which is
comparable with its counterpart in the commutative space, see
Table 1.  Although, the triple gauge boson coupling constants
simultaneously do not vanish the $K_{Z \gamma\gamma}$ is the only
coupling which is appeared in the cross section and it may be
even zero.  Since the values of the triple gauge boson coupling
constants can not be uniquely obtained in the nmNCSM, to be
certain, we may restrict ourselves to the mNCSM where there is not
such a freedom.  In contrast to the nmNCSM in the mNCSM there is
not any triple gauge boson vertex in the electro-weak sector
therefore we have not any diagram at the tree level for the
elastic photon-neutrino scattering.  But in the extended version
of mNCSM which is introduced in section 2 the photon can interact
directly with right handed neutrino therefore at the tree level
there are two Feynman diagrams for the photon-neutrino elastic
scattering which is shown in Fig.2.  The Feynman rule for the
$\gamma\nu\bar\nu $ vertex in the extended mNCSM is given in
 Eq.(\ref{gnn}) as\\
\vspace*{1cm}
\begin{picture}(300,50)(0,0)
\unitlength=1mm \ArrowLine(300,25)(340,9)\ArrowLine(340,9)(380,25)
\Photon(340,-27)(340,9){2}{7.5}
\Vertex(340,9){3}\Text(47,0)[c]{$\Gamma^\mu_{\gamma\nu\bar\nu}=i{\frac
{e}{2}}cos\theta_W(1+\gamma_5)(\theta^{\mu\nu}k_\nu q \!\!\!/ +
\theta^{\rho\mu}q_\rho k \!\!\!/ + \theta^{\nu\rho}k_\nu
q_\rho\gamma^\mu)$.} \nonumber
\end{picture}
\vspace{1cm}\\
Therefore, the invariant amplitude for the first diagram of
Fig.(2) in the center of mass frame can be written as
\begin{eqnarray}
&-i{\cal M}_1&=\varepsilon_{\mu}\varepsilon^{\prime}_\nu
\overline{u}(p^\prime) (ie cos\theta_W)\frac{1}{2}(1+\gamma^5)
[k^\prime.\theta.(k+p)\gamma^\nu \nonumber\\ & &
+\theta^{\nu\alpha}k^{\prime}_\alpha(\rlap/k+\rlap/p)
-\theta^{\nu\alpha}(k_\alpha+p_\alpha)\rlap/k^\prime]
{(-i)(\rlap/k+\rlap/p)\over (k+p)^2}\nonumber\\
& & (-ie cos\theta_W)\frac{1}{2}(1+\gamma^5)
[k.\theta.p\gamma^\mu+k_\beta\theta^{\mu\beta}\rlap/p
-p_\beta\theta^{\mu\beta}\rlap/k]u(p),\nonumber\\
\label{m1}
\end{eqnarray}
which, because of the momentum conservation
$k+p=k^\prime+p^\prime$, the Dirac equations $\rlap/pu(p)=0 ,
\bar{u}(p^\prime)\rlap/p^\prime=0$ and the following identity
\begin{eqnarray}
A_\mu\theta^{\mu\nu}B_\nu=A.\theta.B=
\overrightarrow{\theta}.(\mbox{$\boldmath{A}$}\times\mbox{$\boldmath{B}$}),\label{identity1}
\end{eqnarray}
 where $\overrightarrow{\theta}=(\theta_{23}, \theta_{31}, \theta_{12})$, results in
\begin{eqnarray}
{\cal M}_1=-e^2cos^2\theta_W\varepsilon^{\prime}_\nu
\varepsilon_{\mu}p^\prime_{\alpha}p_{\beta} \theta^{\nu\alpha}
\theta^{\mu\beta}\overline{u}(p^\prime)
\frac{1}{2}(1+\gamma^5)\rlap/k^\prime u(p).
\end{eqnarray}
For the second diagram one similarly has
\begin{eqnarray}
&-i{\cal M}_2&=\varepsilon_{\mu}\varepsilon^{\prime^\ast}_\nu
\overline{u}(p^\prime) (-ie cos\theta_W)\frac{1}{2}(1+\gamma^5)
[k.\theta.(p^\prime-k)\gamma^\mu\nonumber\\ & &
+\theta^{\mu\beta}k_\beta(\rlap/p^\prime-\rlap/k)
-\theta^{\mu\beta}(p_\beta^\prime-k_\beta)\rlap/k]
{-i(\rlap/p^\prime-\rlap/k)\over (p^\prime-k)^2}\nonumber\\ & &
(ie cos\theta_W)\frac{1}{2}(1+\gamma^5)
\left[k^\prime.\theta.p\gamma^\nu+\theta^{\nu\alpha}k^\prime_\alpha\rlap/p
-\theta^{\nu\alpha}p_\alpha\rlap/k^\prime\right]u(p),\nonumber\\
\end{eqnarray}
which after some manipulation yields
\begin{eqnarray} &{\cal
M}_2&=-e^2cos^2\theta_W\varepsilon_{\mu}\varepsilon^{\prime}_\nu
\overline{u}(p^\prime)\frac{1}{2}(1+\gamma^5)\times\nonumber\\ & &
[{(k.\theta.p^\prime)(k^\prime.\theta.p)\over
(p^\prime-k)^2}\gamma^\mu(\rlap/p^\prime- \rlap/k)\gamma^\nu
-p_\alpha
p_\beta^\prime\theta^{\mu\beta}\theta^{\nu\alpha}\rlap/k^\prime\nonumber\\
& & -(k.\theta.p^\prime)(p_\alpha\theta^{\nu\alpha}\gamma^\mu-
p_\alpha^\prime\theta^{\mu\alpha}\gamma^\nu)]u(p).
\end{eqnarray}
Therefore by introducing the appropriate tensor $T^{\mu\nu}$ in
terms of the total amplitude ${\cal M}_{tot}={\cal M}_1+{\cal
M}_2$ one can show that
\begin{equation}
k_{\mu}T^{\mu\nu}=k^\prime_{\nu}T^{\mu\nu}=0.
\end{equation}
  Thus, the
spin-averaged amplitude for $\gamma\nu\rightarrow\gamma\nu$
scattering can be evaluated as
\begin{eqnarray}
&\overline{\mid {\cal M}_{tot}\mid}^2&=\frac{1}{2}e^4cos^4\theta_W
g_{\nu\delta}g_{\mu\eta}\nonumber\\ & &
Tr\left(\rlap/p^\prime\frac{1}{2}(1-\gamma^5)
\sum_{i=1}^{3}I_i^{\mu\nu}\rlap/p
\sum_{j=1}^{3}J_j^{\eta\delta}\right),
\end{eqnarray}
where
\begin{eqnarray}
&&I_1^{\mu\nu}\equiv{(k.\theta.p^\prime)(k^\prime.\theta.p)\over
(p^\prime-k)^2}
\gamma^\mu(\rlap/p^\prime-\rlap/k)\gamma^\nu\ \ ,\nonumber\\
&&I_2^{\mu\nu}\equiv(k.\theta.p^\prime)(p_\alpha\theta^{\nu\alpha}\gamma^\mu-
p_\alpha^\prime\theta^{\mu\alpha}\gamma^\nu)\ \ ,\nonumber\\
&&I_3^{\mu\nu}\equiv(p_\beta p_\alpha^\prime-p_\alpha
p_\beta^\prime)
\theta^{\mu\beta}\theta^{\nu\alpha}\rlap/k^\prime\ \ ,\nonumber\\
&&J_1^{\eta\delta}=I_1^{\delta\eta}\ \ ,\nonumber\\
&&J_2^{\eta\delta}=I_2^{\eta\delta}\ \ ,\nonumber\\
&&J_3^{\eta\delta}=I_3^{\eta\delta}\ \ ,
\end{eqnarray}
which, using the trace theorems, implies
\begin{eqnarray}
&\overline{\mid {\cal M}_{tot}\mid}^2&=\frac{e^4 cos^4\theta_W}{2}
[8\left(k.\theta.p^\prime\right)^4{p.p^\prime\over(p^\prime-k)^2}\nonumber\\
& & +8(k.p)(k^\prime.p)((p^\prime.\theta.\theta.p^\prime)
 (p.\theta.\theta.p)\nonumber\\
& &- \left(p^\prime.\theta.\theta.p\right)^2)
+4\left(k.\theta.p^\prime\right)^2((k^\prime.p)
(p.\theta.\theta.p^\prime\nonumber\\
& &-p.\theta.\theta.p-p^\prime.\theta.\theta.p^\prime)
-3(k.p)\left(p.\theta.\theta.p^\prime\right))].\nonumber \\
\label{square of m}
\end{eqnarray}
To evaluate the total cross section the particle momenta are
shown in Fig.(\ref{com}) and the differential cross section is
given by
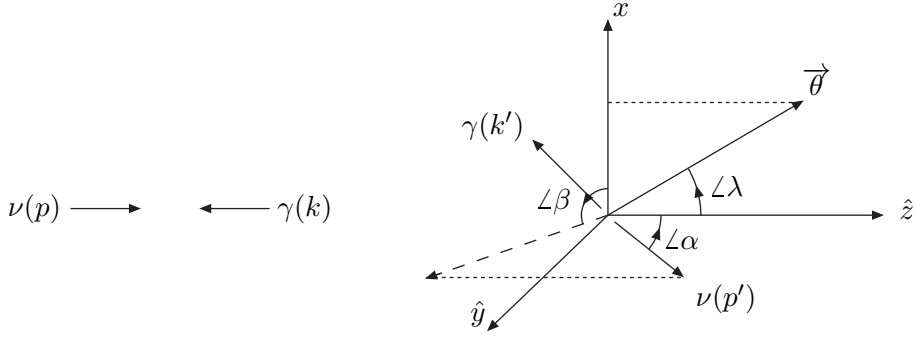
\begin{figure*}[t]
\begin{eqnarray}
\begin{picture}(300,100)(28,-20)
\unitlength=1mm \LongArrow(25,20)(50,20) \LongArrow(100,20)(75,20)
\LongArrow(227.5,17.5)(330,17.5)\Text(120,7)[c]{$\hat{z}$}
\LongArrow(227.5,17.5)(227.5,90)\Text(82,34)[c]{$\hat{x}$}
\LongArrow(227.5,17.5)(183,-25.25)\Text(63,-7)[c]{$\hat{y}$}
\ArrowArc(227.5,17.5)(20,-40,0)\Text(90,3)[c]{$\angle\alpha$}
\LongArrow(227.5,17.5)(300,60)\DashLine(300,60)(227.5,60){1.5}
\ArrowArc(227.5,17.5)(35,0,30)
\Text(108,24)[c]{$\overrightarrow\theta$}
\Text(96,10)[c]{$\angle\lambda$} \LongArrow(225,20)(200,45)
\LongArrow(230,15)(255,-5)
\DashLine(255,-6)(159,-6){1.5}\DashLine(227.5,17.5)(160,-6){5}\LongArrow(165,-4.25)(160,-6)
\ArrowArc(227.5,17.5)(10,90,200)\Text(73,8)[c]{$\angle\beta$}
\Text(4,7)[c]{$\nu(p)$} \Text(40,7)[c]{$\gamma(k)$}
\Text(96,-5)[c]{$\nu(p^\prime)$}
\Text(65,18)[c]{$\gamma(k^\prime)$}
\end{picture}
\nonumber
\end{eqnarray}
\caption{\label{com}The process $\gamma\nu\rightarrow\gamma\nu$ in
the center of mass frame.}
\end{figure*}\\
\begin{eqnarray}
d\sigma=\frac{\overline{\mid {\cal M}_{tot}\mid}^2}
{4\pi^2\times4k.p}\ \frac{d^3\mbox{$\boldmath
P^\prime$}}{2E^\prime_\nu} \ \frac{d^3\mbox{$\boldmath
K^\prime$}}{2E^\prime_\gamma}\ \delta^4(k^\prime+p^\prime-k-p).
\end{eqnarray}
Now by introducing:
\begin{eqnarray}
&&p=(E,\mbox{$\boldmath P$}),\\
&&k=(E,\mbox{$\boldmath -P$}),\\
&&p^\prime=(E^\prime_\nu,\mbox{$\boldmath P$}^\prime),\\
&&k^\prime=(E^\prime_\gamma,\mbox{$\boldmath K$}^\prime),
\end{eqnarray}
the differential cross section can be cast into
\begin{eqnarray}
d\sigma=\frac{\overline{\mid {\cal M}_{tot}\mid}^2}
{4\pi^2\times4k.p}\ \frac{d^3\mbox{$\boldmath
P^\prime$}}{4E^{\prime^2}} \ \delta(2E^\prime-2E),
\end{eqnarray}
where in the center of mass frame $E^\prime\equiv
E^\prime_\gamma=E^\prime_\nu $.  In the relativistic limit
$d^3\mbox{$\boldmath P$}^\prime$ is equal to
$E^{\prime^2}dE^\prime d\beta\ d\cos\alpha$, therefore, in this
limit one has
\begin{eqnarray}
d\sigma=\frac{\overline{\mid {\cal M}_{tot}\mid}^2_{E^\prime=E}}
{2^7\times\pi^2\times k.p} d\beta d\cos\alpha. \label{differential
cross section}
\end{eqnarray}
Now by using the invariant quantities:
\begin{eqnarray}
&&k.p=2E^2,\\
&&p.p^\prime=k.k^\prime=E^2(1-\cos\alpha),\\
&&p.k^\prime=k.p^\prime=E^2(1+\cos\alpha),
\end{eqnarray}
and also the identity given in Eq.(\ref{identity1}) and
\begin{eqnarray}
A_\mu\theta^{\mu\nu}\theta^{\beta}_\nu B_\beta=
A.\theta.\theta.B=\mid\!\!\!\overrightarrow{\theta}\!\!\!\mid^2
(\mbox{$\boldmath{A}$}.\mbox{$\boldmath{B}$})-
(\mbox{$\boldmath{A}$}.\overrightarrow{\theta})
(\mbox{$\boldmath{B}$}.\overrightarrow{\theta}),
\end{eqnarray}
which leads to
\begin{eqnarray}
&&(k.\theta.p^\prime)^2=E^4\mid\!\!\!\overrightarrow\theta\!\!\!\mid^2
\sin^2\!\!\alpha\ \sin^2\!\!\lambda\ \sin^2\!\!\beta\nonumber,\\
&&p.\theta.\theta.p=E^2\mid\!\!\!\overrightarrow{\theta}\!\!\!\mid^2\sin^2\!\!\lambda\nonumber,\\
&&p^\prime.\theta.\theta.p^\prime=E^2\mid\!\!\!\overrightarrow{\theta}\!\!\!\mid^2
(1-\cos^2\!\!\alpha\ \cos^2\!\!\lambda\nonumber\\
&&-\sin^2\!\!\alpha\ \sin^2\!\!\lambda\ \cos^2\!\!\beta
-0.5\sin2\alpha\ \sin2\lambda\ \cos\beta)\nonumber,\\
&&p.\theta.\theta.p^\prime=E^2\mid\!\!\!\overrightarrow{\theta}\!\!\!\mid^2
(\sin^2\!\!\lambda\ \cos\alpha-0.5\sin2\lambda\ \sin\alpha
\cos\beta),\nonumber\\
\end{eqnarray}
one can easily perform the $\beta$ and the $\alpha$ integration
of (\ref{differential cross section}) to find
\begin{equation}
\sigma=0.5\alpha^2cos^4\theta_W{E^6\over\Lambda^8}\label{sigma
final0},
\end{equation}
or
\begin{equation}
\sigma=3.8\times 10^{-32}({M_Z\over\Lambda})^8(\frac{E}{m_e})^6
\,\,\,pb.\label{sigma final}
\end{equation}
By choosing $\Lambda=113 \,\,GeV$ in Eq.(\ref{sigma final}) one
has
\begin{equation}
\sigma=6.7\times 10^{-33}(\frac{E}{m_e})^6 \,\,\,pb,\label{sigma
final1}
\end{equation}
which is equal to the the cross section of photon-neutrino elastic
scattering in the range $m_e\ll E\ll M_W$ in the commutative
standard model given in Eq.(\ref{elastic pn}) while for the cross
section of Eq.(\ref{sigma final1}) there is not such a constraint.
\section{Summary}
In this paper , we extended the non-commutative standard model
based on the minimal $SU(3)\times SU(2)\times U(1)$ gauge group to
include the interaction of the neutral gauge bosons with the
neutrino.  Since in the gauge invariant manner only the particle
with neutral hyper-charge can couple to the hyper-gauge field,
 the right handed neutrino part of the NCSM Lagrangian
density changes as is given in Eq.(\ref{Lagden}). Consequently, we
obtained the Feynman rule for the  $\gamma\nu\bar\nu$-vertex
which does not exist in the minimal extension of the
non-commutative standard model introduced in \cite{NCSM}, while
for the $Z\nu\bar\nu$-vertex we find some corrections given in
Eqs.(\ref{gnn}-\ref{znn}). We explored the photon-neutrino
elastic scattering in both the nmNCSM and the extended version of
mNCSM.  In the former model, the left handed neutrino at the tree
level can interact with photon via Z-gauge boson exchange as is
shown in Fig.(1). We showed that the cross section grows as $E^6$
in the center of mass and depends on the new undetermined
constant, $K_{Z \gamma\gamma}$, as well as the parameter of
non-commutativity, see Eq.(\ref{cross-nm}). The cross section for
$K_{Z \gamma\gamma}=0.1$ varies in the range $10^{-42}$-$10^{-46}$
$cm^2$ for $\Lambda\sim 100-1000 GeV$ and $E\sim 0.1M_Z$ which is
comparable with its counterpart in the commutative space though
$K_{Z \gamma\gamma}$ varies in the range $-0.3\leq K_{Z
\gamma\gamma}\leq 0.1$ and it may be zero. Nevertheless, the
photon-neutrino elastic scattering is also examined in the
extended version of mNCSM where photon can interact directly with
neutrino.  In this case there are two Feynman diagrams at the
tree level which are presented in Fig.(2).  Since the parameter of
non-commutativity is the only mass scale, the cross section should
be proportional to $\alpha^2\Lambda^{-8}E^6$ which is explicitly
obtained in Eq.(\ref{sigma final0}).  Comparison of
Eq.(\ref{sigma final1}) and Eq.(\ref{elastic pn}) with
Eq.(\ref{inelastic pn}) shows that the three cross sections are
equal for $E=6.5 GeV$ while the value of the photon-neutrino
elastic scattering cross section in the non-commutative space at
$E=10 GeV$ is about two times the value of its counterpart in the
commutative space.  Therefore, at sufficiently high energies the
process $\nu\gamma\rightarrow\nu\gamma$ in the non-commutative
space dominates the processes $\nu\gamma\rightarrow\nu\gamma$ and
$\nu\gamma\rightarrow\nu\gamma\gamma$ in the commutative space.
Nonetheless, for the higher values of $\Lambda$ the elastic cross
section in the NC-space will be comparable with the elastic one
in the commutative space at the higher energies.  For example for
$\Lambda= 1000 GeV$ at $E=500 GeV$ it is still one percent of the
cross section of $\nu\gamma\rightarrow\nu\gamma$ in the SM while
they are equal at $E\sim 1000GeV $.  Therefore,  in the high
energy limit the right handed neutrino has the same contribution
to the photon-neutrino scattering as the left handed one and is
not the expectator particle.
\begin{table}
\caption{ The total cross section for
$\gamma\nu\rightarrow\gamma\nu$ in the nmNCSM given in
Eq.(\ref{cross-nm}) for $K_{Z \gamma\gamma}=0.1$, the mNCSM given
in Eq.(\ref{sigma final0}) and in the standard model (SM) obtained
in \cite{abbas}. }
\begin{center}
\begin{tabular}{|c|c|c|c|}\hline
$\sigma(\nu\gamma\rightarrow\nu\gamma)$ & nmNCSM  & mNCSM  & SM\\
$(cm^2)$& ($\Lambda\sim 100-1000 GeV$) & ($\Lambda\sim 100-1000
GeV$) &
\\ \hline
$E=1 MeV$ &$3.4\times 10^{-67}-3.4\times 10^{-71}$&$1\times 10^{-66} -1\times 10^{-74} $&$4\times 10^{-67} $\\
\hline $E=10 GeV$&$3.4\times 10^{-43}-3.4\times 10^{-47}$
&$1\times 10^{-42}-1\times 10^{-50} $&
$2\times 10^{-43} $\\
\hline
\end{tabular}
\end{center}
\end{table}
\section*{Acknowledgement}
 The financial support of IUT research council is acknowledged.

\end{document}